\documentclass[aps,twocolumn,showpacs,preprintnumbers,noshowpacs,6pt]{revtex4-1}
\usepackage[T2A]{fontenc}
\usepackage[utf8]{inputenc}
\usepackage[english,russian]{babel} 
\usepackage{hyperref}
\usepackage{mathtools}
\usepackage{graphicx}
\usepackage{amsmath}
\usepackage[tablename=Table]{caption}
\usepackage{subcaption}
\usepackage{amssymb}
\usepackage{pgfplots}
\pgfplotsset{width=10cm,compat=1.9}
\usepackage{color}
\usepackage{array}
\usepackage{dsfont}
\usepackage{tikz}
\usepackage{braket}
\usepackage{feynmf}
\usepackage{threeparttable}
\usepackage{indentfirst}
\usepackage{siunitx}
\usepackage{bm}
\usepackage{wrapfig}
\usepackage{float}
\usepackage{geometry}
\usepackage{bm}
\usepackage{etoolbox}
\patchcmd{\thebibliography}{\section*}{\section}{}{}
\usepackage{multirow}
\usepackage{lipsum,booktabs}
\usepackage{breqn}

\makeatother

\begin{document}

\title{Relativistic calculations of the energies of the low-lying $1sns$, $1snp$, $1snd$ states and the probabilities of the one-photon $1snl\to 1sn'l'$ transitions in heliumlike uranium}

\author{N. K. Dulaev$^{1,2, *}$, M. Y. Kaygorodov$^{1}$, A. V. Malyshev$^{1}$, I. I. Tupitsyn$^{1}$, V. M. Shabaev$^{1,2}$}
\affiliation{${}^{1}$Department of Physics$,$ St$.$ Petersburg State University$,$ Universitetskaya 7$/$9$,$ 199034 St$.$ Petersburg$,$ Russia\\
${}^{2}$Petersburg Nuclear Physics Institute named by B$.$P$.$ Konstantinov of National Research Center “Kurchatov Institute”$,$ Orlova roscha 1$,$ 188300 Gatchina$,$ Leningrad region$,$ Russia\\
${}^*$e-mail: st069071@student.spbu.ru}

\begin{abstract}

For heliumlike uranium, the energies of the singly-excited $1sns$, $1snp$, and $1snd$ states with $n\leq 4$ and the probabilities of the one-photon $1s3d\to 1s2p$, $1s3p\to 1s2s$, $1s3p\to 1s2p$ and $1s4d\to 1s2p$ transitions are evaluated. 
The calculations are performed within the Breit approximation using the configuration-interaction method in the basis of the Dirac-Fock-Sturm orbitals.
The QED corrections to the energy levels are calculated employing the model-QED-operator approach.
The nuclear recoil, frequency-dependent Breit-interaction, nuclear polarization, and nuclear deformation corrections are taken into account as well.

\end{abstract}

\maketitle
%
%
%
%
%
%
\section{Introduction}

The study of highly charged ions plays an important role in modern physics \cite{Shabaev2006, Volotka2013, Kozlov_2018, Shabaev_2018, 2018_SafronovaM_RMP90, Indelicato_2019}. 
The comparison of the various properties of highly charged ions measured in high-precision experiments with the results of theoretical calculations makes it possible to test the methods of quantum electrodynamics (QED) at the strong-coupling regime, to improve the accuracy of the fundamental constants and nuclear-structure parameters.
The investigation of highly charged ions with two electrons~--- heliumlike ions~--- is of particular interest since they are the simplest atomic systems in which the interelectronic-interaction effects are manifested. 
\par

During the last decades, the experimental accuracy of the
transition-energy measurements in highly charge ions has been significantly
improved. For instance, the uncertainty of the Lamb-shift measurement in H-like
uranium constitutes $2\%$~\cite{UraniumLambShift2, uraniumLambShift1}. 
Even better precision has been achieved in experiments with Li-like ions \cite{Schweppe1991, Brandau2003, Beiersdorfer2005, Brandau2008}. 
The high-precision measurements of the transition energies in He-like ions were performed in a number of works \cite{Beiersforfer1989, Trassinelli2007, Trassinelli2009, Trassinelli2011, Rudolph2013, Trassinelli2013, Crespo2014, Crespo2015, Beiersdorfer2015, Machado2018}.
To analyze the experimental results, the theoretical evaluation of the transition energies as well as the transition probabilities, is required. 
The most precise calculations of the transition energies in He-like ions were
performed in Refs.~\cite{ArtemyevQED, MalyshevQED2019, KozhedubUranium, YerokhinUranium}. The transition probabilities were evaluated in
Refs.~\cite{TrPr1, TrPr2, TrPr3, TrPr4}.
\par
Currently, one of the experiments to study the electronic structure of highly charged ions is carried out at the storage ring CRYRING@ESR at the GSI in Darmstadt~\cite{Lestinsky2016, 2022_PfaffleinP_PS97}.
Within this experiment, the investigation of the x-ray spectrum associated with the transitions from the states with single excitations into the~$L$,~$M$, and ~$N$ shells to the ground state is planned for He-like uranium.
Typical energies of these transitions are tens of keV.
In addition to the transition energies, the measurements of the population of the excited states in He-like uranium are scheduled.
To correctly identify the levels and to analyze the experimental data, the theoretical calculations of the electronic-structure parameters of the singly-excited states of~$\mathrm{U}^{90+}$ are in demand.

\par
The present work aims at the theoretical study of the electronic structure of He-like uranium.
The energy levels of singly-excited $1sns$, $1snp$, $1snd$ states with $n \leq 4$ are calculated within the Breit approximation by means of the configuration-interaction method in the basis of the Dirac-Fock-Sturm orbitals.
The QED corrections to the energy levels are evaluated using the model-QED-operator approach.
The nuclear recoil, frequency-dependent Breit interaction, nuclear polarization, and nuclear deformation corrections are taken into account as well.
In addition, for the $1s3d\to 1s2p$, $1s3p\to 1s2s$, $1s3p\to 1s2p$, $1s4d\to 1s2p$ transitions the one-photon-transition probabilities are calculated.

\par
The paper is organized as follows.
In Sec. II, the theoretical methods used for the calculations are described. 
The numerical results as well as the comparison with other works are discussed in Sec. III.

\par
The atomic units are used throughout the paper.

\section{Methods of calculations}
The calculations of the energies and transition probabilities in heliumlike uranium are performed within the Breit approximation by means of the configuration-interaction method in the basis of the Dirac-Fock-Sturm orbitals (CI-DFS)~\cite{RelIsTup, 2005_TupitsynI_PhysRevA, Tup2018}.
To describe the $N$-electron system (in our case $N=2$) within the Breit approximation, the Dirac-Coulomb-Breit Hamiltonian (DCB) is used 
\begin{equation}
    \hat{H}^{\textrm{DCB}} = \Lambda^{(+)} \left[ \hat{H}_{\textrm{D}} + \hat{V}_{\textrm{C}} + \hat{V}_{\textrm{B}} \right] \Lambda^{(+)}.
    \label{DCB}
\end{equation}
Here $\hat{H}_{\textrm{D}}$ is the sum of the one-electron Dirac Hamiltonians
\begin{equation}
    \hat{H}_{\textrm{D}} = \sum_i^N \hat{h}^{\textrm{D}}_i,
\end{equation}
\begin{equation}
    \hat{h}^{\textrm{D}}_i = c(\pmb{\alpha}_i  \cdot \pmb{p}_i ) + mc^2(\beta-1) + V_{\textrm{nucl}}(r_i),
    \label{dir}
\end{equation}
where $\pmb{p}$ is the momentum operator, $\pmb{\alpha}$ and $\beta$ are the Dirac matrices, and $V_{\textrm{nucl}}$ is the potential of the nucleus.
In the calculations, the Fermi nuclear-charge-distribution model is employed and 
the root-mean-square radius for the uranium nucleus is taken from Ref.~\cite{Angeli}.
The operator $\hat{V}_{\textrm{C}}$ is the sum of the two-electron Coulomb-interaction operators
\begin{equation}
    \hat{V}_{\textrm{C}} = \frac{1}{2} \sum_{i\neq j}^N  \frac{1}{r_{ij}}, \quad r_{ij} = |\pmb{r}_i - \pmb{r}_j|,
\end{equation}
the operator $\hat{V}_{\textrm{B}}$ is the sum of the Breit-interaction operators
\begin{equation}
    \hat{V}_{\textrm{B}} = -\frac{1}{2} \sum_{i \neq j}^N \frac{1}{2r_{ij}}\Big[\bm{\alpha}_i\cdot\bm{\alpha}_j+\frac{(\bm{\alpha}_i\cdot\bm{r}_{ij})(\bm{\alpha}_j\cdot\bm{r}_{ij})}{r_{ij}^2}\Big]. 
    \label{breit}
\end{equation}
In the Hamiltonian (\ref{DCB}), $\Lambda^{(+)}$ is the product of the one-electron projectors on the positive-energy eigenvalues of the Dirac-Fock (DF) operator.
\par
In the CI-DFS method, the many-electron wave function $\Psi(J M_J)$ with the total angular momentum $J$ and its projection $M_J$ is expanded in the basis of the configuration-state functions (CSFs) $\Phi_\alpha(J M_J)$,
\begin{equation}
     \Psi( J M_J) = \sum_\alpha C_\alpha(J M_J) \Phi_\alpha(J M_J).
\end{equation}
The CSFs are the eigenfunctions of the operator $\hat{J}^2$.
They are constructed as the appropriate linear combinations of the Slater determinants. 
The mixing coefficients $C_\alpha(J M_J) $ are determined from the solution of the matrix eigenvalue problem 
\begin{equation}\label{eq:ci_problem}
    H^{\mathrm{DCB}} C (J M_J)  = E^{\mathrm{DCB}}(J) C(J M_J),
\end{equation}
where $H^{\mathrm{DCB}}$ is the Hamiltonian matrix in the basis of the CSFs and $C(JM_J)$ is the vector of the mixing coefficients. 
\par
The one-electron basis is constructed as follows. For the occupied $nl$ states and for the low-lying virtual $n'l'$ states, where $n' \leq n$ and $l' \leq l$, the orbitals are obtained as numerical solutions of the DF equations. 
All other virtual orbitals correspond to the solutions of the DF equations in the finite basis set of the Sturmian functions.
The Sturmian functions are the numerical solutions of the Dirac-Fock-Sturm equation 
\begin{equation}
 (\hat{h}^{\textrm{DF}} - \varepsilon_0) \phi_j = \lambda_j W(r) \phi_j,
\end{equation}
where $\hat{h}^{\textrm{DF}}$ is the DF Hamiltonian, $\varepsilon_0$ is the one-electron reference energy of the occupied DF $ns$, $np$ or $nd$ orbital, and $W(r)$ is the weight function
\begin{equation}
    W(r) = \left[  \frac{1-e^{-(ar)^2}}{(ar)^2} \right]^n.
\end{equation}
The parameters $a$ and $n$ are adjusted to achieve the fastest convergence of the energy $E(J)$ with respect to the number of the virtual orbitals.

\par
The QED corrections are calculated using the model-QED-operator approach~\cite{ShabaevModelQED, QEDMODshabaev1, QEDMODshabaev2}. 
The model-QED operator~$\hat{V}^{\mathrm{Q}}$ is constructed in the way to reproduce the exact values of the diagonal and off-diagonal matrix elements of the one-loop QED contributions for the low-lying states of H-like ions.
The practical application of the model-QED operator~$\hat{V}^{\mathrm{Q}}$ consists in adding it to the DCB Hamiltonian~$\hat{H}^{\textrm{DCB}}$,
\begin{equation}\label{eq:h_dcbq}
    \hat{H}^{\textrm{DCBQ}} = \Lambda^{(+)}_{\mathrm{Q}} \left[ \hat{H}^{\textrm{D}} + \hat{V}^{\textrm{C}} + \hat{V}^{\textrm{B}} + \hat{V}^{\textrm{Q}}\right] \Lambda^{(+)}_{\mathrm{Q}},
\end{equation}
and then finding the lowest eigenvalues of the matrix~$H^{\mathrm{DCBQ}}$
\begin{equation}\label{eq:ci_problem_qed}
    H^{\mathrm{DCBQ}} C (J M)  = E^{\mathrm{DCBQ}}(J) C(J M).
\end{equation}
We should note that the operator $V^{\mathrm{Q}}$ is included into the Hamiltonian~$\hat{h}^{\mathrm{DF}}$ at the basis-construction stage, therefore, the projectors $\Lambda^{(+)}_{\mathrm{Q}}$ in Eq. (\ref{eq:h_dcbq}) differ from the projectors $\Lambda^{(+)}$ in Eq. (\ref{DCB}).
The QED correction to the energy of a level is determined as the difference of the total energies
\begin{equation}\label{eq:delta_e_qed}
    \Delta E^{\mathrm{QED}}(J) = E^{\mathrm{DCBQ}}(J) - E^{\mathrm{DCB}}(J).
\end{equation}
The described procedure allows one to partially take into account the screened QED corrections within the multi-configuration calculations.
\par

The nuclear recoil effect, caused by the finite mass of the nucleus, leads to the shift of the energy levels.
Within the lowest-order relativistic approximation and to first order in the electron-to-nucleus mass ratio $m/M$, the nuclear-recoil Hamiltonian can be written as~\cite{ShabaevTeor1985_Eng, ShabaevY1988_Eng, Palmer_1987, ShabaevMC2}
\begin{equation}
    \hat{H}^{\textrm{MS}} = \frac{1}{2M}\sum_{i,j} \left[ \pmb{p}_i \cdot \pmb{p}_j - \frac{\alpha Z}{r_i} \left( \pmb{\alpha}_i + \frac{(\pmb{\alpha}_i \cdot \pmb{r}_i)\pmb{r}_i}{r_i^2} \right) \cdot \pmb{p}_j \right]. \label{MS}
\end{equation}
The QED corrections to the nuclear recoil effect were calculated earlier (see, e.g., Refs.~\cite{ShabaevTeor1985_Eng, ShabaevY1988_Eng, Artemyev_1995_2, Artemyev_1995_1, Shabaev1998_2, Malyshev_2018_recoil, Anisimova2022} and references therein). 
In the present paper, the nuclear recoil correction to the energy level, $\Delta E^{\mathrm{MS}}$, is defined as the sum of the expectation value of the operator $\hat{H}^{\textrm{MS}}$, evaluated using the correlated many-electron function $\Psi(JM_J)$, and the corresponding one-electron QED corrections.

\par
The frequency-dependent Breit-interaction correction to the energy level, $\Delta E^{\mathrm{FB}}$, is calculated as follows. 
Let us consider the one-photon-exchange operator
    \begin{equation}
        I(\omega) = \alpha_1^{\mu} \alpha_2^{\nu} D_{\mu \nu} (\omega, \pmb{r}_{12}),
    \end{equation}
where $D_{\mu \nu}$ is the photon propagator in the Coulomb gauge
\begin{equation}\label{oo}
\begin{split}
    D_{00}(\omega, \pmb{r}_{12}) =& \frac{1}{r_{12}}, \quad D_{i0}=D_{0i}=0, \quad i=1,2,3,\\
    D_{il}(\omega, \pmb{r}_{12}) =&  4\pi \int \frac{d \pmb{k}}{(2 \pi)^3} \frac{\textrm{exp}(i\pmb{k}\cdot \pmb{r}_{12})}{\omega^2-\pmb{k}^2+i0} \left( \delta_{il} - \frac{k_i k_l}{\pmb{k}^2} \right), \\
    i,l=1,2,3. 
\end{split} 
\end{equation}
Considering the $\omega\to 0$ limit in Eq. (\ref{oo}), we obtain the standard form of the Breit interaction (\ref{breit}).
The correction $\Delta E^{\mathrm{FB}}$ is evaluated as the expectation value of the symmetrized one-photon-exchange operator \cite{Mittleman1972, Shabaev_1993, shabaevgreen, Tupitsyn2020_Eng, Tupitsyn2022_Eng} with the wave functions obtained by the CI-DFS method.

The nuclear polarization and nuclear deformation corrections to the energy levels of He-like uranium are calculated according to Refs.~\cite{Plunien1991, Plunien1995, NEFIODOV1996, Kozhedub2008, KozhedubUranium}.

Let us consider the probability of the transition of the many-electron system from the state \textbf{$\beta$} with the total angular momentum $J_{\beta}$ to the state $\alpha$ with the total angular momentum $J_{\alpha}$. 
The probability of spontaneous emission of a photon with the frequency $\omega$ and the multipolarity $\lambda L$ ($\lambda = E$ for the electric-type transitions and $\lambda =M$ for the magnetic type transitions) is given by the expression \cite{Grant_1974}

\begin{equation}
    A_L^{(\lambda)}(\beta, \alpha) = 2 \alpha \omega \frac{2L +1}{2 J_{\beta} +1}\left|\braket{\alpha || T^{(\lambda)}_{L} ||\beta} \right|^2,
\end{equation}
where $\braket{\alpha || T^{(\lambda)}_{L} ||\beta}$ is the reduced matrix element of the multipole transition operator $T^{(\lambda)}_{L}$. 
To calculate the transition probabilities, the many-electron functions $\Psi_\alpha$ and $\Psi_\beta$ obtained by means of the CI-DFS method are used. 
The many-electron functions are evaluated for the DCB Hamiltonian with the model-QED operator included (\ref{eq:h_dcbq}). 
Therefore, the calculated transition probabilities partially incorporate the QED corrections.

\section{Numerical results and discussion}

For heliumlike uranium, the systematic calculations of the energies of the $1sns$, $1snp$, $1snd$ states with $n\leq 4$ are carried out. When constructing the many-electron basis within the CI-DFS method, all the possible single and double excitations from the reference configuration, which corresponds to the occupied state, into a space spanned by a given number of virtual orbitals are considered.
The orbitals with $n\leq 17$ for each quantum number $l\leq 11$ are used as a one-electron basis set in the calculations of the total energy $E^{\textrm{DCB}}$: the total number of the one-electron functions is $138$. 
For each considered state, the uncertainty associated with the incompleteness of the one-electron basis is determined from the analysis of the convergence of the total energy $E^{\textrm{DCB}}$ with respect to the number of the one-electron functions.
It is established that the uncertainty of $E^{\textrm{DCB}}$ due to the finite size of the basis does not exceed a value of the order of $0.1$ eV for all the considered states.

\par

Further, various corrections to the energies $E^{\textrm{DCB}}$ are calculated.
The QED corrections are evaluated using the model-QED-operator approach according to Eqs. (\ref{eq:h_dcbq}) --- (\ref{eq:delta_e_qed}). When calculating the QED correction, a significantly smaller basis of the one-electron functions is used, since the correction $\Delta E^{\mathrm{QED}}$ converges with respect to the number of the one-electron functions faster than the total energy $E^{\textrm{DCB}}$. 
The orbitals with $n\leq 13$ for each quantum number $l\leq 4$ are used in the QED-correction calculations, the total number of the used functions is $55$. 
The uncertainty of the evaluated QED corrections associated with the incompleteness of the basis set constitutes approximately $0.01$ eV for the all considered states. 
Additionally, the nuclear recoil correction, $\Delta E^{\textrm{MS}}$, and the frequency-dependent Breit interaction correction, $\Delta E^{\textrm{FB}}$, are calculated. 
These corrections are evaluated using even smaller number of the virtual orbitals, however, the numerical accuracy for these corrections is several orders of magnitude higher than the accuracy for the energy $E^{\textrm{DCB}}$.

\par
In Table~\ref{table1new} for the ground state of heliumlike uranium, the values of the energy obtained using the DCB Hamiltonian, $E^{\mathrm{DCB}}$, the QED correction, $\Delta E^{\mathrm{QED}}$, the nuclear recoil correction, $\Delta E^{\mathrm{MS}}$, the frequency-dependent Breit-interaction correction, $\Delta E^{\mathrm{FB}}$, the nuclear polarization and deformation corrections, $\Delta E^{\mathrm{PD}}$, as well as the total energy including all the corrections, $E^{\mathrm{tot}}$, are presented. 
The obtained results are compared with the data from Ref.~\cite{KozhedubUranium}. 
In Ref.~\cite{KozhedubUranium}, the rigorous calculations of the one-electron (self-energy and vacuum polarization) and screened QED corrections, the two-photon-exchange contribution, and also the higher-order interelectronic-interaction contributions within the Breit approximation were performed.
Moreover, the contributions of the one-electron two-loop diagrams were taken into account there. 
The calculations of the nuclear recoil effect in both the present work and Ref.~\cite{KozhedubUranium} have been carried out taking into account the corresponding QED contribution.

\par
The DCB energy obtained in the present work is compared with the DCB energy from Ref.~\cite{KozhedubUranium} calculated using the projectors~$\Lambda^{(+)}$ in the Eq. ~(\ref{DCB}) which correspond to the Dirac equation (\ref{dir}). 
The value of the QED corrections calculated in the present work is compared with the sum of the one-electron and screened QED corrections calculated in Ref.~\cite{KozhedubUranium} using the local Dirac-Fock potential (LDF).
The frequency-dependent Breit-interaction correction is compared with the value of the one-photon-exchange contribution from Ref.~\cite{KozhedubUranium}. 
Our total energy is compared with the total value from Ref.~\cite{KozhedubUranium}, which also includes the nuclear polarization and deformation corrections.
\begin{table}[H]
    \centering
    \caption{The ground-state energy of heliumlike uranium calculated using the Dirac-Coulomb-Breit Hamiltonian, $E^{\textrm{DCB}}$, and various corrections to this value: the frequency-dependent Breit interaction correction, $\Delta E^{\mathrm{FB}}$, the QED correction, $\Delta E^{\mathrm{QED}}$, the nuclear recoil correction, $\Delta E^{\mathrm{MS}}$, the nuclear polarization and deformation corrections, $\Delta E^{\mathrm{PD}}$. 
    The value $E^{\textrm{tot}}_{1s1s}$ is the total energy (eV). 
    The total energy is compared with the result of Ref.~\cite{KozhedubUranium}.}
    \begin{tabular}{
    l|
    S[table-format=8.2]
    }
    \hline
    \multicolumn{1}{c|}{Contribution}      & \multicolumn{1}{c}{Value}    \\ \hline
    $E^{\textrm{DCB}}$         & -261910.84 \\
    $\Delta E^{\textrm{FB}}$  & -0.02   \\
    $\Delta E^{\textrm{QED}}$       & 527.00     \\
    $\Delta E^{\textrm{MS}}$      & 0.92       \\
    $\Delta E^{\textrm{PD}}$      & -0.62       \\
    $E^{\textrm{tot}}_{1s1s}$         & -261383.56 \\
    $E^{\textrm{tot}}_{1s1s}$ \cite{KozhedubUranium} & -261386.15
    \end{tabular}
    \label{table1new}
\end{table}
\par

The comparison shows that the results of the present calculations for the ground state of heliumlike uranium are in agreement with the result of Ref.~\cite{KozhedubUranium}.
Indeed, the DCB energy of the ground state of heliumlike uranium, calculated in the present work, is $-261910.84$ eV, while in Ref.~\cite{KozhedubUranium} this quantity equals $-261910.73$ eV, that is within our estimated numerical uncertainty of $0.1$ eV.
The correction $\Delta E^{\mathrm{FB}}$ in Ref.~\cite{KozhedubUranium} is strictly zero, since the one-photon exchange between the $1s$ electrons occurs at the zero frequency of the virtual photon, $\omega=0$. 
However, in the present work the value of $\Delta E^{\mathrm{FB}}$ deviates from zero due to the mixing of the states which have different energies. 
For the ground state of heliumlike uranium, the QED correction calculated in the present work by means of the model-QED-operator method is $527.00$ eV, which agrees up to the $0.5\%$ uncertainty with the sum of the one-electron and screened QED corrections, $523.01$ eV, obtained in Ref.~\cite{KozhedubUranium}. 
The value of the nuclear recoil correction, $\Delta E^{\mathrm{MS}}$, equals $0.92$ eV, and it is in good agreement with the result of $0.93$ eV obtained in Ref.~\cite{KozhedubUranium}.
The difference of the total ground-state energies for heliumlike uranium $E^{\textrm{tot}}_{1s1s}$ obtained in the present work and Ref.~\cite{KozhedubUranium} is about $2.5$ eV, and it is mainly due to the lack of an accurate consideration of the two-photon-exchange contribution, the approximate treatment of the screened QED contributions, and partial taking into account the contribution of the two-loop diagrams in the present work.

\par

In Table~\ref{table2new} for the excited $(1s2s)_0$, $(1s2s)_1$, $(1s2p_{1/2})_0$, $(1s2p_{3/2})_2$, $(1s2p_{1/2})_1$, $(1s2p_{3/2})_1$ states of heliumlike uranium, the results of the calculations of the energies obtained using the DCB Hamiltonian, $E^{\mathrm{DCB}}$, the QED corrections, $\Delta E^{\mathrm{QED}}$, the nuclear recoil corrections, $\Delta E^{\mathrm{MS}}$, the frequency-dependent Breit-interaction corrections, $\Delta E^{\mathrm{FB}}$, the nuclear polarization and deformation corrections, $\Delta E^{\mathrm{PD}}$, are presented. 
The total energies, $E^{\mathrm{tot}}$, which include all the corrections, and energies relative to the ground state, $E^{\mathrm{tot}} - E^{\mathrm{tot}}_{1s1s}$, are given. 
The latter results are compared with the related values from Ref.~\cite{KozhedubUranium}.
\onecolumngrid\
\begin{table}[H]
\centering
\caption{The energies of the $1s2s$ and $1s2p$ states of heliumlike uranium calculated using the Dirac-Coulomb-Breit Hamiltonian, $E^{\textrm{DCB}}$, and various corrections to these values: the frequency-dependent Breit-interaction corrections, $\Delta E^{\mathrm{FB}}$, the QED corrections, $\Delta E^{\mathrm{QED}}$, the nuclear recoil corrections, $\Delta E^{\mathrm{MS}}$, the nuclear polarization and deformation corrections, $\Delta E^{\mathrm{PD}}$.
$E^{\textrm{tot}}_{1s1s}$ are the total energies and $E^{\mathrm{tot}} - E^{\textrm{tot}}_{1s1s}$ are the total energies relative to the ground state (eV).
The latter values are compared with the total results from Ref.~\cite{KozhedubUranium}.}
\label{table2new}
\begin{tabular}{c|
l|
S[table-format=10.5]|
c|
l|
S[table-format=10.5]
}
\hline
\multicolumn{1}{c|}{State}                    & \multicolumn{1}{c|}{Contribution}                         & \multicolumn{1}{c|}{Value}   & \multicolumn{1}{c|}{State}                    & \multicolumn{1}{c|}{Contribution}                         & \multicolumn{1}{c}{Value}    \\ \hline
\multirow{8}{*}{$(1s2s)_0$}    & $E^{\textrm{DCB}}$                                    & -165418.06 & \multirow{8}{*}{$(1s2p_{3/2})_2$} & $E^{\textrm{DCB}}$                                    & -161115.78 \\
                            & $\Delta E^{\textrm{FB}}$                               & 0.67       &                             & $\Delta E^{\textrm{FB}}$                               & -7.05      \\
                            & $\Delta E^{\textrm{QED}}$                                    & 314.79     &                             & $\Delta E^{\textrm{QED}}$                                    & 275.05     \\
                            & $\Delta E^{\textrm{MS}}$                                 & 0.58       &                             & $\Delta E^{\textrm{MS}}$                                 & 0.50       \\
                            & $\Delta E^{\textrm{PD}}$                                 & -0.37        &                               &  $\Delta E^{\textrm{PD}}$                            & -0.31 \\
                            & $E^{\textrm{tot}}$                                  & -165102.39 &                             & $E^{\textrm{tot}}$                                    & -160847.60 \\
                            & $E^{\textrm{tot}} - E^{\textrm{tot}}_{1s1s}$         & 96281.17   &                             & $E^{\textrm{tot}} - E^{\textrm{tot}}_{1s1s}$        & 100535.96  \\
                            &$E^{\textrm{tot}} - E^{\textrm{tot}}_{1s1s}$ \cite{KozhedubUranium} & 96281.78(54)   &                             & $E^{\textrm{tot}} - E^{\textrm{tot}}_{1s1s}$ \cite{KozhedubUranium} & 100536.95(54)   \\ \hline 
\multirow{8}{*}{$(1s2s)_1$}    & $E^{\textrm{DCB}}$                                    & -165673.15 & \multirow{8}{*}{$(1s2p_{1/2})_1$} & $E^{\textrm{DCB}}$                                    & -165488.45 \\
                            & $\Delta E^{\textrm{FB}}$                               & 0.23       &                             & $\Delta E^{\textrm{FB}}$                               & 0.10       \\
                            & $\Delta E^{\textrm{QED}}$                                    & 315.89     &                             & $\Delta E^{\textrm{QED}}$                                    & 272.92     \\
                            & $\Delta E^{\textrm{MS}}$                                 & 0.58       &                             & $\Delta E^{\textrm{MS}}$                                 & 0.54       \\
                            & $\Delta E^{\textrm{PD}}$                                 & -0.37        &                               &  $\Delta E^{\textrm{PD}}$                            & -0.32 \\
                            & $E^{\textrm{tot}}$                                   & -165356.82 &                             & $E^{\textrm{tot}}$                                   & -165215.20 \\
                            & $E^{\textrm{tot}} - E^{\textrm{tot}}_{1s1s}$        & 96026.74   &                             & $E^{\textrm{tot}} - E^{\textrm{tot}}_{1s1s}$         & 96168.36   \\
                            & $E^{\textrm{tot}} - E^{\textrm{tot}}_{1s1s}$ \cite{KozhedubUranium} & 96027.07(54)   &                             & $E^{\textrm{tot}} - E^{\textrm{tot}}_{1s1s}$ \cite{KozhedubUranium} & 96169.43(54)    \\ \hline 
\multirow{8}{*}{$(1s2p_{1/2})_0$} & $E^{\textrm{DCB}}$                                    & -165379.04 & \multirow{8}{*}{$(1s2p_{3/2})_1$} & $E^{\textrm{DCB}}$                                    & -161052.09 \\
                            & $\Delta E^{\textrm{FB}}$                               & 0.32       &                             & $\Delta E^{\textrm{FB}}$                               & 2.96       \\
                            & $\Delta E^{\textrm{QED}}$                                    & 272.73     &                             & $\Delta E^{\textrm{QED}}$                                    & 275.16     \\
                            & $\Delta E^{\textrm{MS}}$                                 & 0.53       &                             & $\Delta E^{\textrm{MS}}$                                 & 0.54       \\
                            & $\Delta E^{\textrm{PD}}$                                 & -0.32        &                               &  $\Delta E^{\textrm{PD}}$                            & -0.31 \\
                            & $E^{\textrm{tot}}$                                  & -165105.77 &                             & $E^{\textrm{tot}}$                                   & -160773.74 \\
                            & $E^{\textrm{tot}} - E^{\textrm{tot}}_{1s1s}$         & 96277.79   &                             & $E^{\textrm{tot}} - E^{\textrm{tot}}_{1s1s}$       & 100609.82  \\
                            & $E^{\textrm{tot}} - E^{\textrm{tot}}_{1s1s}$ \cite{KozhedubUranium} & 96279.01(54)    &                             & $E^{\textrm{tot}} - E^{\textrm{tot}}_{1s1s}$ \cite{KozhedubUranium}& 100610.68(54) 
\end{tabular}
\end{table}
\twocolumngrid\
\par

Table~\ref{table2new} shows that the results of the present work are in reasonable agreement with the results of Ref.~\cite{KozhedubUranium}. 
The energies $E^{\mathrm{DCB}}$ obtained in the present work are consistent with the results of Ref.~\cite{KozhedubUranium} within the uncertainty of $0.1$ eV estimated for the Dirac-Coulomb-Breit equation solutions.
For various excited states, as in the case of the ground state, the difference of the contributions $\Delta E^{\mathrm{QED}}$ between this work and Ref.~\cite{KozhedubUranium} is about $0.5\%$. 
For the total energies $E^{\mathrm{tot}}$, the difference between the final results does not exceed $2.5$ eV. 
The systematic deviation decreases if we consider the transition energy to the ground state~--- the difference between the results becomes about $1$ eV. 
The reasons of these deviations are the same as for the ground-state values: in the present work, the QED corrections are taken into account approximately and the two-photon-exchange contribution beyond the Breit approximation is excluded from the consideration.
\par
\onecolumngrid\
\begin{table}[H]
\centering
\caption{The energies of the $1sns$ and $1snp$ states with $n=3,4$ of heliumlike uranium calculated using the Dirac-Coulomb-Breit Hamiltonian, $E^{\textrm{DCB}}$, and various corrections to these values: the frequency-dependent Breit-interaction corrections, $\Delta E^{\mathrm{FB}}$, the QED corrections, $\Delta E^{\mathrm{QED}}$, the nuclear recoil corrections, $\Delta E^{\mathrm{MS}}$, the nuclear polarization and deformation corrections, $\Delta E^{\mathrm{PD}}$. $E^{\textrm{tot}}_{1s1s}$ are the total energies and $E^{\mathrm{tot}} - E^{\textrm{tot}}_{1s1s}$ are the total energies relative to the ground state (eV).
The latter values are compared with the total result from Ref.~\cite{YerokhinUranium}.}
\label{table3new}
\begin{tabular}{c|
l|
S[table-format=8.2]|
S[table-format=8.2]|
S[table-format=8.2]|
S[table-format=8.2]|
S[table-format=8.2]|
S[table-format=8.2]}
\hline
\multicolumn{1}{c|}{$n$} & \multicolumn{1}{ c|}{Contribution}                     &\multicolumn{1}{c|}{$(1sns)_0$} & \multicolumn{1}{c|}{$(1sns)_1$}  &\multicolumn{1}{c|}{$(1snp_{1/2})_0$}  & \multicolumn{1}{c|}{$(1snp_{3/2})_2$}  & \multicolumn{1}{c|}{$(1snp_{1/2})_1$}  &\multicolumn{1}{c}{$(1snp_{3/2})_1$}  \\ 
\hline
\multirow{8}{*}{$n=3$}  & $E^{\textrm{DCB}}$                                                        & -146389.84 & -146456.53 & -146380.77 & -145104.14 & -146408.38 & -145084.29 \\
                      & $\Delta E^{\textrm{FB}}$                                                    & 0.24       & 0.08       & 0.55       & -2.06      & 0.18       & 0.92       \\
                      & $\Delta E^{\textrm{QED}}$                                                   & 281.11     & 281.34     & 269.13     & 269.52     & 269.08     & 269.54     \\
                      & $\Delta E^{\textrm{MS}}$                                                    & 0.51       & 0.51       & 0.50       & 0.49       & 0.50      & 0.50       \\
                      & $\Delta E^{\textrm{PD}}$                                                    & -0.33       & -0.33       & -0.31       & -0.31       & -0.31       & -0.31       \\
                      & $E^{\textrm{tot}}$                                                          & -146108.30 & -146174.94 & -146110.90 & -144836.51 & -146138.93 & -144813.64 \\
                      & $E^{\textrm{tot}} - E^{\textrm{tot}}_{1s1s}$                                & 115275.26 &	115208.62	 & 115272.66	& 116547.05	& 115244.63	& 116569.92 \\
                      & $E^{\textrm{tot}} - E^{\textrm{tot}}_{1s1s}$ \cite{YerokhinUranium}         & 115276.70  & 115209.77  & 115273.83  & 116548.37  & 115245.92  & 116571.21  \\ 
                      \hline
\multirow{8}{*}{$n=4$}  & $E^{\textrm{DCB}}$                                                        & -139923.01 & -139949.50 & -139919.59 & 	-139387.14 & 	-139930.42	& -139378.70 \\
                      & $\Delta E^{\textrm{FB}}$                                                    & 0.10       & 0.03       & 0.28       & -0.86      & 0.09       & 0.39       \\
                      & $\Delta E^{\textrm{QED}}$                                                   & 272.64     & 272.71     & 267.71     & 267.93     & 267.73     & 267.95     \\
                      & $\Delta E^{\textrm{MS}}$                                                    & 0.49       & 0.49       & 0.48       & 0.48       & 0.48       & 0.48       \\
                      & $\Delta E^{\textrm{PD}}$                                                    & -0.31       & -0.31       & -0.31       & -0.31       & -0.31       & -0.31       \\
                      & $E^{\textrm{tot}}$                                                          & -139650.09 & -139676.58  & -139651.43 &	-139119.89 &	-139662.43 &	-139110.19\\
                      & $E^{\textrm{tot}} - E^{\textrm{tot}}_{1s1s}$                                & 121733.47 &	121706.98  &	121732.13 &	122263.67 &	121721.13 &	122273.37 \\
                      & $E^{\textrm{tot}} - E^{\textrm{tot}}_{1s1s}$ \cite{YerokhinUranium}         & 121734.83  & 121708.20  & 121733.31  & 122264.98  & 121722.35  & 122274.65   
\end{tabular}
\label{tableNew2}
\end{table}
\twocolumngrid\
In Table~\ref{table3new}, the results for the energies of the $(1sns)_0$, $(1sns)_1$, $(1snp_{1/2})_0$, $(1snp_{1/2})_1$, $(1snp_{3/2})_1$, $(1snp_{3/2})_2$  states with $n=3,4$ are presented. 
For each state, the values of the individual contributions, the total binding energy, and the energy relative to the ground state are given. The transitions energies to the ground state are compared with the results of Ref.~\cite{YerokhinUranium}. 
In Ref.~\cite{YerokhinUranium}, the interelectronic interaction was treated within the Breit approximation using the configuration-interaction method.
Additionally, in Ref.~\cite{YerokhinUranium} the frequency-dependent Breit interaction and the nuclear recoil corrections were taken into account. 
The consideration of the one-loop QED corrections was based in Ref.~\cite{YerokhinUranium} on the employed by us model-QED-operator approach.
Furthermore, the two-loop QED corrections were considered there. 
In Ref.~\cite{YerokhinUranium}, the uncertainty of the theoretical calculations of the transition energies was estimated to be at the level of $1$ eV.
From Table~\ref{table3new} it can be seen that for the transitions considered, our results are in reasonable agreement with the results of Ref.~\cite{YerokhinUranium}: the differences for all the transition energies do not exceed 2 eV.

\par
Finally, in Table~\ref{tableNew3} the results for the energies of the $(1snd_{3/2})_1$, $(1snd_{3/2})_2$, $(1snd_{5/2})_2$, $(1snd_{5/2})_3$ states with $n=3,4$ are shown.
For each state, the values of the individual contributions, the total energy, and the energy of the singly-excited state relative to the ground one are presented. 
Based on the comparison of the present results for the transition energies to the ground state from the $1sns$, $1snp$ states with $n=1,2$ and $1sns$, $1snp$ states with $n=3,4$ with the theoretical prediction Ref.~\cite{KozhedubUranium} and Ref.~\cite{YerokhinUranium}, respectively, we estimate the uncertainty of the obtained results for the corresponding transitions from the $1snd$ states with $n=3,4$ to be at the level of $2$ eV. 
This uncertainty includes the error due to the QED effects beyond the model-QED-operator approach and the error due to the uncertainty of the root-mean-square radius of the $^{238}\mathrm{U}$ nucleus.
\onecolumngrid\
\begin{table}[H]
\centering
\caption{The energies of the $1snd$ states with $n\leq4$ of heliumlike uranium calculated using the Dirac-Coulomb-Breit Hamiltonian, $E^{\textrm{DCB}}$, and various corrections to these values: the frequency-dependent Breit-interaction corrections, $\Delta E^{\mathrm{FB}}$, the QED corrections, $\Delta E^{\mathrm{QED}}$, the nuclear recoil corrections, $\Delta E^{\mathrm{MS}}$, the nuclear polarization and deformation corrections, $\Delta E^{\mathrm{PD}}$. $E^{\textrm{tot}}_{1s1s}$ are the total energies and $E^{\mathrm{tot}} - E^{\textrm{tot}}_{1s1s}$ are the total energies relative to the ground state (eV).}
\begin{tabular}{c|
l|
S[table-format=8.2]|
S[table-format=8.2]|
S[table-format=8.2]|
S[table-format=8.2]
}
\hline
\multicolumn{1}{c|}{$n$}                     & \multicolumn{1}{c|}{Contribution}       & \multicolumn{1}{c|}{$(1snd_{3/2})_2$} & \multicolumn{1}{c|}{$(1snd_{3/2})_1$} & \multicolumn{1}{c|}{$(1snd_{5/2})_3$} & \multicolumn{1}{c}{$(1snd_{5/2})_2$} \\ 
\hline
\multirow{7}{*}{$n=3$} & $E^{\textrm{DCB}}$                                    & -145092.76                    & -145084.47                    & -144759.69                    & -144754.50                    \\
                     & $\Delta E^{\textrm{FB}}$                                & 0.02                          & 0.12                          & 0.03                          & 0.02                          \\
                     & $\Delta E^{\textrm{QED}}$                               & 266.48                        & 266.49                        & 267.20                        & 267.21                        \\
                     & $\Delta E^{\textrm{MS}}$                                & 0.49                          & 0.49                          & 0.49                          & 0.49                         \\
                     & $\Delta E^{\textrm{PD}}$                                & -0.31                          & -0.31                          & -0.31                          & -0.31                          \\
                     & $E^{\textrm{tot}}$                                      & -144826.07                    & -144817.69                    & -144492.28                    & -144487.08                    \\ 
                     & $E^{\textrm{tot}} - E^{\mathrm{tot}}_{1s1s}$           & 116557.49                      & 116565.87                     & 116891.28                     & 116896.48 \\
                     \hline
\multirow{7}{*}{$n=4$} & $E^{\textrm{DCB}}$                                   & -139382.25                      &	-139378.74	               & -139240.45	                 & -139238.15                    \\
                     & $\Delta E^{\textrm{FB}}$                               & 0.01                           & 0.07                          & 0.02                          & 0.00                          \\
                     & $\Delta E^{\textrm{QED}}$                              & 266.62                        & 266.63                        & 267.04                        & 266.92                        \\
                     & $\Delta E^{\textrm{MS}}$                               & 0.48                          & 0.48                          & 0.48                          & 0.48                          \\
                     & $\Delta E^{\textrm{PD}}$                               & -0.31                          & -0.31                          & -0.31                          & -0.31                          \\
                     & $E^{\textrm{tot}}$                                     & -139115.44                    &	-139111.87                   &	-138973.22                    &	-138971.06                \\ 
                     & $E^{\textrm{tot}} - E^{\mathrm{tot}}_{1s1s}$           & 122268.12                    & 122271.69                     & 122410.34                     & 122412.50 
\end{tabular}
\label{tableNew3}
\end{table}
\twocolumngrid\
\par
\begin{table}[H]
    \centering
    
    \caption{The probabilities of the one-photon transitions with the lowest possible multipolarity $\lambda L$, $A_L^{(\lambda)}(\beta, \alpha)$ (s$^{-1}$), and the transition energies, $\Delta E_{\beta\alpha}$ (eV), for heliumlike uranium. 
    The $1s$ orbital is omitted in the designations of the initial and final states for the sake of brevity.
    }
    \begin{tabular}{
    c|
    c|
    S[table-format=7.3]|
    c
    }
    \hline
    \multicolumn{1}{c|}{Transition   $\beta \to  \alpha$}                            &  \multicolumn{1}{c|}{$\lambda L$} & \multicolumn{1}{c|}{$\Delta E_{\beta\alpha}$} & \multicolumn{1}{c}{$A_L^{(\lambda)}(\beta, \alpha)$}  \\ 
    \hline
    $ (3d_{3/2})_2 \rightarrow (2p_{3/2})_1 $ & E1   & 15947.7                & $ 0.747\cdot 10^{14} $            \\ 
    $ (3d_{3/2})_2 \rightarrow (2p_{3/2})_2 $ & E1   & 16021.5                & $ 0.645\cdot 10^{15} $            \\
    $ (3d_{3/2})_1 \rightarrow (2p_{3/2})_2 $ & E1   & 16029.9                & $ 0.119\cdot 10^{15} $            \\
    $ (3d_{5/2})_3 \rightarrow (2p_{3/2})_1 $ & M2   & 16281.5                & $ 0.149\cdot 10^{12} $            \\
    $ (3d_{5/2})_2 \rightarrow (2p_{3/2})_1 $ & E1   & 16286.7                & $ 0.390\cdot 10^{16} $            \\
    $ (3d_{5/2})_3 \rightarrow (2p_{3/2})_2 $ & E1   & 16355.3                & $ 0.436\cdot 10^{16} $            \\
    $ (3d_{5/2})_2 \rightarrow (2p_{3/2})_2 $ & E1   & 16360.5                & $ 0.437\cdot 10^{15} $            \\
    $ (3d_{3/2})_1 \rightarrow (2p_{1/2})_0 $ & E1   & 20288.1                & $ 0.294\cdot 10^{16} $            \\ 
    $ (3p_{3/2})_1 \rightarrow (2s_{1/2})_0 $ & E1   & 20288.8                & $ 0.776\cdot 10^{15} $            \\
    $ (3p_{3/2})_1 \rightarrow (2p_{1/2})_0 $ & M1   & 20292.1                & $ 0.224\cdot 10^{12} $            \\ 
    $ (3d_{3/2})_2 \rightarrow (2p_{1/2})_1 $ & E1   & 20389.1                & $ 0.443\cdot 10^{16} $            \\ 
    $ (3d_{3/2})_1 \rightarrow (2p_{1/2})_1 $ & E1   & 20397.5                & $ 0.148\cdot 10^{16} $            \\ 
    $ (3p_{3/2})_2 \rightarrow (2s_{1/2})_1 $ & E1   & 20520.3                & $ 0.114\cdot 10^{16} $            \\ 
    $ (3p_{3/2})_1 \rightarrow (2s_{1/2})_1 $ & E1   & 20543.2                & $ 0.373\cdot 10^{15} $            \\ 
    $ (3d_{5/2})_2 \rightarrow (2p_{1/2})_0 $ & M2   & 20618.7                & $ 0.215\cdot 10^{12} $            \\ 
    $ (4d_{5/2})_2 \rightarrow (2p_{3/2})_1 $ & E1   & 21802.7                & $ 0.127\cdot 10^{16} $            \\ 
    $ (4d_{5/2})_3 \rightarrow (2p_{3/2})_2 $ & E1   & 21874.4                & $ 0.141\cdot 10^{16} $            
    \end{tabular}
    \label{t9}
    \end{table} 

In Table \ref{t9}, the one-photon transition probabilities for the $1s3d\to 1s2p$, $1s3p\to 1s2s$, $1s3p\to 1s2p$, and $1s4d\to 1s2p$ transitions with the lowest possible multipolarities  are presented. 
The transition energies are obtained from the results given in Tables~\ref{table1new}~--- \ref{tableNew3}.
In the calculations of the transition probabilities, the orbitals with $n\leq14$ for $l=0$, $n\leq11$ for $l=1$ and $n\leq9$ for $l=3,4$ are used, the total number of the one-electron functions is $37$.
The dipole $E1$ $(1s3d_{5/2})_3\to (1s2p_{3/2})_2$ and $(1s3d_{3/2})_2\to (1s2p_{1/2})_1$ transitions have the largest probabilities, which are approximately $0.44\cdot 10^{16}$ s$^{-1}$. 
The energies of these transitions are equal to $16360.5$ eV and $20389.1$ eV, respectively.

\section{Summary}
In the present work, the energies of the $1sns$, $1snp$, $1snd$ states with $n\leq 4$ of heliumlike uranium are calculated using the configuration-interaction method in the basis of the Dirac-Fock-Sturm orbitals. 
The energies and probabilities of the one-photon $1s 3d\to 1s 2p$, $1s3p\to 1s 2s$, $1s3p\to 1s 2p$, and $1s4d\to 1s2p$ transitions with the lowest possible multipolarities are evaluated. 
The QED corrections to the energies of the states are taken into account using the model-QED-operator approach. 
In addition, the nuclear recoil corrections, the frequency-dependent Breit-interaction corrections, and the nuclear polarization and deformation corrections are calculated to the energies of the states and transition energies.

\subsection*{Acknowledgment}
This work was supported by the Russian Science Foundation (Grant № 22-62-00004, https://rscf.ru/project/22-62-00004/).

\bibliographystyle{OS2.bst}
\bibliography{main.bib}

\begin{thebibliography}{10}

\bibitem{Shabaev2006}
V.~Shabaev, O.~Andreev, A.~Artemyev, S.~Baturin, A.~Elizarov, Y.~Kozhedub,
  N.~Oreshkina, I.~Tupitsyn, V.~Yerokhin, O.~Zherebtsov.
\newblock Int. J. Mass Spectrometry, {\bf 251},~109 (2006).

\bibitem{Volotka2013}
A.V. Volotka, D.A. Glazov, G.~Plunien, V.M. Shabaev.
\newblock Ann. Phys., {\bf 525},~636 (2013).

\bibitem{Kozlov_2018}
M.G. Kozlov, M.S. Safronova, J.R.C. {L{\'o}pez-Urrutia}, P.O. Schmidt.
\newblock Rev. Mod. Phys, {\bf 90},~045005 (2018).

\bibitem{Shabaev_2018}
V.M. Shabaev, A.I. Bondarev, D.A. Glazov, M.Y. Kaygorodov, Y.S. Kozhedub, I.A.
  Maltsev, A.V. Malyshev, R.V. Popov, I.I. Tupitsyn, N.A. Zubova.
\newblock Hyperfine Interact., {\bf 239},~1 (2018).

\bibitem{2018_SafronovaM_RMP90}
M.S. Safronova, D.~Budker, D.~DeMille, D.F.J. Kimball, A.~Derevianko, C.W.
  Clark.
\newblock Rev. Mod. Phys, {\bf 90},~025008 (2018).

\bibitem{Indelicato_2019}
P.~Indelicato.
\newblock J. Phys. B, {\bf 52},~232001 (2019).

\bibitem{UraniumLambShift2}
T.~St{\"o}hlker, P.H. Mokler, F.~Bosch, R.W. Dunford, F.~Franzke, O.~Klepper,
  C.~Kozhuharov, T.~Ludziejewski, F.~Nolden, H.~Reich, P.~Rymuza, Z.~Stachura,
  M.~Steck, P.~Swiat, A.~Warczak.
\newblock Phys. Rev. Lett., {\bf 85},~3109 (2000).

\bibitem{uraniumLambShift1}
A.~Gumberidze, T.~St{\"o}hlker, D.~Bana{\'s}, K.~Beckert, P.~Beller, H.F.
  Beyer, F.~Bosch, S.~Hagmann, C.~Kozhuharov, D.~Liesen, F.~Nolden, X.~Ma, P.H.
  Mokler, M.~Steck, D.~Sierpowski, S.~Tashenov.
\newblock Phys. Rev. Lett., {\bf 94},~223001 (2005).

\bibitem{Schweppe1991}
J.~Schweppe, A.~Belkacem, L.~Blumenfeld, N.~Claytor, B.~Feinberg, H.~Gould,
  V.E. Kostroun, L.~Levy, S.~Misawa, J.R. Mowat, M.H. Prior.
\newblock Phys. Rev. Lett., {\bf 66} (11),~1434 (1991).

\bibitem{Brandau2003}
C.~Brandau, C.~Kozhuharov, A.~M\"uller, W.~Shi, S.~Schippers, T.~Bartsch,
  S.~B\"ohm, C.~B\"ohme, A.~Hoffknecht, H.~Knopp, N.~Gr\"un, W.~Scheid,
  T.~Steih, F.~Bosch, B.~Franzke, P.H. Mokler, F.~Nolden, M.~Steck,
  T.~St\"ohlker, Z.~Stachura.
\newblock Phys. Rev. Lett., {\bf 91} (7),~073202 (2003).

\bibitem{Beiersdorfer2005}
P.~Beiersdorfer, H.~Chen, D.B. Thorn, E.~Tr\"abert.
\newblock Phys. Rev. Lett., {\bf 95} (23),~233003 (2005).

\bibitem{Brandau2008}
C.~Brandau, C.~Kozhuharov, Z.~Harman, A.~M\"uller, S.~Schippers, Y.S. Kozhedub,
  D.~Bernhardt, S.~B\"ohm, J.~Jacobi, E.W. Schmidt, P.H. Mokler, F.~Bosch, H.J.
  Kluge, T.~St\"ohlker, K.~Beckert, P.~Beller, F.~Nolden, M.~Steck,
  A.~Gumberidze, R.~Reuschl, U.~Spillmann, F.J. Currell, I.I. Tupitsyn, V.M.
  Shabaev, U.D. Jentschura, C.H. Keitel, A.~Wolf, Z.~Stachura.
\newblock Phys. Rev. Lett., {\bf 100} (7),~073201 (2008).

\bibitem{Beiersforfer1989}
P.~Beiersdorfer, M.~Bitter, S.~von Goeler, K.W. Hill.
\newblock Phys. Rev. A, {\bf 40} (1),~150 (1989).

\bibitem{Trassinelli2007}
M.~Trassinelli, S.~Boucard, D.S. Covita, D.~Gotta, A.~Hirtl, P.~Indelicato,
  {\relax {\'E}.-O}.~Le~Bigot, J.M.F. dos Santos, L.M. Simons, L.~Stingelin,
  J.F.C.A. Veloso, A.~Wasser, J.~Zmeskal.
\newblock Journal of Physics: Conference Series, {\bf 58},~129 (2007).

\bibitem{Trassinelli2009}
M.~Trassinelli, A.~Kumar, H.F. Beyer, P.~Indelicato, R.~Märtin, R.~Reuschl,
  Y.S. Kozhedub, C.~Brandau, H.~Bräuning, S.~Geyer, A.~Gumberidze, S.~Hess,
  P.~Jagodzinski, C.~Kozhuharov, D.~Liesen, U.~Spillmann, S.~Trotsenko,
  G.~Weber, D.F.A. Winters, T.~Stöhlker.
\newblock Europhys. Lett., {\bf 87},~63001 (2009).

\bibitem{Trassinelli2011}
M.~Trassinelli, A.~Kumar, H.F. Beyer, P.~Indelicato, R.~Märtin, R.~Reuschl,
  Y.S. Kozhedub, C.~Brandau, H.~Bräuning, S.~Geyer, A.~Gumberidze, S.~Hess,
  P.~Jagodzinski, C.~Kozhuharov, D.~Liesen, U.~Spillmann, S.~Trotsenko,
  G.~Weber, D.F.A. Winters, T.~Stöhlker.
\newblock Phys. Scr., {\bf 2011},~014003 (2011).

\bibitem{Rudolph2013}
J.K. Rudolph, S.~Bernitt, S.W. Epp, R.~Steinbr\"ugge, C.~Beilmann, G.V. Brown,
  S.~Eberle, A.~Graf, Z.~Harman, N.~Hell, M.~Leutenegger, A.~M\"uller,
  K.~Schlage, H.C. Wille, H.~Yava\ifmmode~\mbox{\c{s}}\else \c{s}\fi{},
  J.~Ullrich, J.R. Crespo L\'opez-Urrutia.
\newblock Phys. Rev. Lett., {\bf 111} (10),~103002 (2013).

\bibitem{Trassinelli2013}
S.~Schlesser, S.~Boucard, D.S. Covita, J.M.F. dos Santos, H.~Fuhrmann,
  D.~Gotta, A.~Gruber, M.~Hennebach, A.~Hirtl, P.~Indelicato, {\relax
  {\'E}.-O}.~Le~Bigot, L.M. Simons, L.~Stingelin, M.~Trassinelli, J.F.C.A.
  Veloso, A.~Wasser, J.~Zmeskal.
\newblock Phys. Rev. A, {\bf 88} (2),~022503 (2013).

\bibitem{Crespo2014}
K.~Kubi\ifmmode~\check{c}\else \v{c}\fi{}ek, P.H. Mokler, V.~M\"ackel,
  J.~Ullrich, J.R.C. L\'opez-Urrutia.
\newblock Phys. Rev. A, {\bf 90} (3),~032508 (2014).

\bibitem{Crespo2015}
S.W. Epp, R.~Steinbr\"ugge, S.~Bernitt, J.K. Rudolph, C.~Beilmann, H.~Bekker,
  A.~M\"uller, O.O. Versolato, H.C. Wille, H.~Yava\ifmmode~\mbox{\c{s}}\else
  \c{s}\fi{}, J.~Ullrich, J.R. Crespo L\'opez-Urrutia.
\newblock Phys. Rev. A, {\bf 92} (2),~020502 (2015).

\bibitem{Beiersdorfer2015}
P.~Beiersdorfer, G.V. Brown.
\newblock Phys. Rev. A, {\bf 91} (3),~032514 (2015).

\bibitem{Machado2018}
J.~Machado, C.I. Szabo, J.P. Santos, P.~Amaro, M.~Guerra, A.~Gumberidze,
  G.~Bian, J.M. Isac, P.~Indelicato.
\newblock Phys. Rev. A, {\bf 97} (3),~032517 (2018).

\bibitem{ArtemyevQED}
A.N. Artemyev, V.M. Shabaev, V.A. Yerokhin, G.~Plunien, G.~Soff.
\newblock Phys. Rev. A, {\bf 71} (6),~062104 (2005).

\bibitem{MalyshevQED2019}
A.V. Malyshev, Y.S. Kozhedub, D.A. Glazov, I.I. Tupitsyn, V.M. Shabaev.
\newblock Phys. Rev. A, {\bf 99} (1),~010501 (2019).

\bibitem{KozhedubUranium}
Y.S. Kozhedub, A.V. Malyshev, D.A. Glazov, V.M. Shabaev, I.I. Tupitsyn.
\newblock Phys. Rev. A, {\bf 100} (6),~062506 (2019).

\bibitem{YerokhinUranium}
V.A. Yerokhin, A.~Surzhykov.
\newblock J. Phys. Chem. Ref. Data, {\bf 48},~033104 (2019).

\bibitem{TrPr1}
E.~Lindroth, S.~Salomonson.
\newblock Phys. Rev. A, {\bf 41},~4659 (1990).

\bibitem{TrPr2}
P.~Indelicato.
\newblock Phys. Rev. Lett., {\bf 77},~3323 (1996).

\bibitem{TrPr3}
P.~Indelicato, V.M. Shabaev, A.V. Volotka.
\newblock Phys. Rev. A, {\bf 69},~062506 (2004).

\bibitem{TrPr4}
W.L. Wiese, J.R. Fuhr.
\newblock J. Phys. Chem. Ref. Data, {\bf 38},~565 (2009).

\bibitem{Lestinsky2016}
M.~Lestinsky, V.~Andrianov, B.~Aurand \emph{et~al.}.
\newblock The European Physical Journal Special Topics, {\bf 225},~797 (2016).

\bibitem{2022_PfaffleinP_PS97}
P.~Pf{\"a}fflein, S.~Allgeier, S.~Bernitt, A.~Fleischmann, M.~Friedrich,
  C.~Hahn, D.~Hengstler, M.O. Herdrich, A.~Kalinin, F.M. Kr{\"o}ger, P.~Kuntz,
  M.~Lestinsky, B.~L{\"o}her, E.B. Menz, T.~Over, U.~Spillmann, G.~Weber,
  B.~Zhu, C.~Enss, T.~St{\"o}hlker.
\newblock Phys. Scr., {\bf 97},~114005 (2022).

\bibitem{RelIsTup}
I.I. Tupitsyn, V.M. Shabaev, J.R. {Crespo L{\'o}pez-Urrutia}, I.~Dragani{\'c},
  R.~Soria~Orts, J.~Ullrich.
\newblock Phys. Rev. A, {\bf 68},~022511 (2003).

\bibitem{2005_TupitsynI_PhysRevA}
I.I. Tupitsyn, A.V. Volotka, D.A. Glazov, V.M. Shabaev, G.~Plunien, J.R.
  {Crespo L{\'o}pez-Urrutia}, A.~Lapierre, J.~Ullrich.
\newblock Phys. Rev. A, {\bf 72},~062503 (2005).

\bibitem{Tup2018}
I.I. Tupitsyn, N.A. Zubova, V.M. Shabaev, G.~Plunien, T.~St\"ohlker.
\newblock Phys. Rev. A, {\bf 98} (2),~022517 (2018).

\bibitem{Angeli}
I.~Angeli, K.~Marinova.
\newblock At. Data. Nucl. Data Tables, {\bf 99},~69 (2013).

\bibitem{ShabaevModelQED}
V.M. Shabaev, I.I. Tupitsyn, V.A. Yerokhin.
\newblock Phys. Rev. A, {\bf 88} (1),~012513 (2013).

\bibitem{QEDMODshabaev1}
V.M. Shabaev, I.I. Tupitsyn, V.A. Yerokhin.
\newblock Comp. Phys. Commun., {\bf 189},~175 (2015).

\bibitem{QEDMODshabaev2}
V.M. Shabaev, I.I. Tupitsyn, V.A. Yerokhin.
\newblock Comp. Phys. Commun., {\bf 223},~69 (2018).

\bibitem{ShabaevTeor1985_Eng}
V.M. Shabaev.
\newblock Theor. Math. Phys., {\bf 63},~588 (1985).

\bibitem{ShabaevY1988_Eng}
V.M. Shabaev.
\newblock Sov. J. Nucl. Phys., {\bf 47},~69 (1988).

\bibitem{Palmer_1987}
C.W.P. Palmer.
\newblock J. Phys. B, {\bf 20},~5987 (1987).

\bibitem{ShabaevMC2}
V.M. Shabaev.
\newblock Phys. Rev. A, {\bf 57},~59 (1998).

\bibitem{Artemyev_1995_2}
A.N. Artemyev, V.M. Shabaev, V.A. Yerokhin.
\newblock Phys. Rev. A, {\bf 52} (3),~1884 (1995).

\bibitem{Artemyev_1995_1}
A.N. Artemyev, V.M. Shabaev, V.A. Yerokhin.
\newblock J. Phys. B, {\bf 28},~5201 (1995).

\bibitem{Shabaev1998_2}
V.M. Shabaev, A.N. Artemyev, T.~Beier, G.~Plunien, V.A. Yerokhin, G.~Soff.
\newblock Phys. Rev. A, {\bf 57} (6),~4235 (1998).

\bibitem{Malyshev_2018_recoil}
A.V. Malyshev, R.V. Popov, V.M. Shabaev, N.A. Zubova.
\newblock J. Phys. B, {\bf 51},~085001 (2018).

\bibitem{Anisimova2022}
I.S. Anisimova, A.V. Malyshev, D.A. Glazov, M.Y. Kaygorodov, Y.S. Kozhedub,
  G.~Plunien, V.M. Shabaev.
\newblock Phys. Rev. A, {\bf 106} (6),~062823 (2022).

\bibitem{Mittleman1972}
M.H. Mittleman.
\newblock Phys. Rev. A, {\bf 5} (6),~2395 (1972).

\bibitem{Shabaev_1993}
V.M. Shabaev.
\newblock J. Phys. B, {\bf 26},~4703 (1993).

\bibitem{shabaevgreen}
V.M. Shabaev.
\newblock Phys. Rep., {\bf 356},~119 (2002).

\bibitem{Tupitsyn2020_Eng}
I.I. Tupitsyn, S.V. Bezborodov, A.V. Malyshev, D.V. Mironova, V.M. Shabaev.
\newblock Opt. Spektrosk., {\bf 128} (1),~24 (2020).

\bibitem{Tupitsyn2022_Eng}
S.V. Bezborodov, I.I. Tupitsyn, A.V. Malyshev, D.V. Mironova, V.M. Shabaev.
\newblock Opt. Spektrosk., {\bf 130} (10),~1471 (2022).

\bibitem{Plunien1991}
G.~Plunien, B.~M\"uller, W.~Greiner, G.~Soff.
\newblock Phys. Rev. A, {\bf 43} (11),~5853 (1991).

\bibitem{Plunien1995}
G.~Plunien, G.~Soff.
\newblock Phys. Rev. A, {\bf 51} (2),~1119 (1995).

\bibitem{NEFIODOV1996}
A.V. Nefiodov, L.N. Labzowsky, G.~Plunien, G.~Soff.
\newblock Phys. Lett. A, {\bf 222},~227 (1996).

\bibitem{Kozhedub2008}
Y.S. Kozhedub, O.V. Andreev, V.M. Shabaev, I.I. Tupitsyn, C.~Brandau,
  C.~Kozhuharov, G.~Plunien, T.~St\"ohlker.
\newblock Phys. Rev. A, {\bf 77} (3),~032501 (2008).

\bibitem{Grant_1974}
I.P. Grant.
\newblock J. Phys. B, {\bf 7},~1458 (1974).

\end{thebibliography}
\end{document}